\documentstyle[psbox,fleqn]{WORKSHOP}

\newcommand{\vb}{\vec b}
\newcommand{\vg}{\vec g}

\newcommand{\vtau}{\vec\tau}
\newcommand{\td}{T_{\rm burst}}
\newcommand{\tb}{T_{\rm bk}}
\newcommand{\ndet}{N_{\rm det}}
\newcommand{\ldet}{L_{\rm det}}
\newcommand{\nmask}{N_{\rm mask}}
\newcommand{\lmask}{L_{\rm mask}}

\newcommand{\topt}{t_{\rm opt}}

\begin{document}
\title{Determination of X-Ray Transient Source Positions\\
By Bayesian Analysis of Coded Aperture Data}
%
\author{
Carlo Graziani,$^1$ Donald Q. Lamb,$^1$ and Raphael Slawinski$^2$
\\[12pt]  
%
$^1$  Dept. of Astronomy \& Astrophysics, University of Chicago \\
$^2$  Dept. of Geophysics, University of Calgary \\
{\it E-mail: c-graziani@uchicago.edu} 
}

\abst{
We present a new method of transient point source deconvolution for
coded-aperture X-Ray detectors.  Our method is based upon the
calculation of the likelihood function and its interpretation as a
probability density for the transient source position by an application
of Bayes' Theorem.  The method obtains point estimates of source
positions by finding the maximum of this probability density, and
interval estimates of prescribed probability by choosing suitable
contours of constant probability density.  We give the results of
simulations that we performed to test the method.  We also derive
approximate analytic expressions for the predicted performance of the
method.  These estimates underline the intuitively plausible properties
of the method and provide a sound quantitative basis for the design of
coded-aperture systems.
}

\kword{techniques: image processing --- gamma rays: bursts}
\maketitle

\section{Introduction}

The nature and origin of gamma-ray bursts is a mystery of twenty-five
years' standing.  The identification of counterparts at other
wavelengths is the ``Holy Grail'' of GRB studies, since it is widely
believed that this is the most likely approach to crack the mystery. 
The consensus is that the next step in GRB instrumentation is to obtain
few-arc-minute or better scale GRB sky locations in bulk, possibly in
real time.  This can be accomplished effectively by means of a
triggered coded-aperture X-ray instrument such as the one planned for
HETE (Ricker 1997).

Several processing methods for coded-aperture data have been proposed,
including cross correlation (Fenimore 1978), least-squares fitting
(Doty 1978), and Maximum Entropy (Sims et al. 1980, Willingdale et al.
1984).  Skinner and Nottingham (1993) have described a
maximum-likelihood fitting technique.  We present here a Bayesian scheme
for analyzing coded-aperture data from such transient events.  The
method is based on the calculation of the joint likelihood function for
two stretches of data:  the stretch covering the transient event
itself, and a stretch before and/or afterwards, which provides
information about the background.  We interpret the likelihood thus
obtained as a posterior probability density for the transient event
location by an application of Bayes' Theorem (for a lucid discussion of
astrophysical applications of Bayesian inference, see Loredo 1992).

This probability density has several uses.  Its maximum provides a
point estimate for the location of the transient source.  We can also
use it to obtain a 68\% credible region for the location of the
source.  Finally, we can use semi-analytical approximations to this
probability distribution to predict the performance of the method, and
to state detector design criteria useful for optimizing the angular
resolution of the instrument.

\section{The Posterior Probability Density}

We begin by exhibiting the posterior probability density for the
transient position.  The discussion in this section is rather similar
to the discussion in Loredo (1992) of Bayesian inference of a Poisson
mean, which may usefully be read for comparison.

We assume that a (one- or two-dimensional) coded-aperture instrument is
illuminated by a transient source.  The instrument consists of a
position-sensitive detector with $\ndet$ position bins, beneath a
coded-aperture mask.  We denote the ``background'' counts observed
during a period $\tb$ prior to the onset of the transient event by
$\vb$, where the $i$th component of $\vb$ is $b_i$,
$i=1,\ldots,\ndet$.  We denote by $\vg$ the gross counts observed
during a time $\td$ while the transient event was occurring. The
background $\vb$ includes the diffuse X-ray background and the particle
background, as well as any steady point sources in the field of view. 
The gross counts $\vg$ reflect the shadow pattern of the coded-aperture
mask on the detector as cast by the illumination of the transient
source, superposed upon the previously measured steady background.

We denote by $\Omega\equiv(\theta,\phi)$ the direction towards the
transient source.  We denote the characteristic shadow pattern of the
mask illuminated by a source in the direction $\Omega$ by
$\vtau(\Omega)$, where we adopt the normalization
$\sum_{i=1}^{\ndet}\tau_i=1$.  We further denote the expected number of
total counts due to the transient by $\omega$, so that the expected
number of counts in the $i$th bin due to the transient source is
$\omega\tau_i$.

If we assume prior probability densities that are uniform in the
background Poisson rates, in $d\omega$, and in
$d\Omega=d\cos\theta\,d\phi$, and integrate over the unknown background
rates and over $\omega$, we obtain the following expression for the
desired posterior probability density:
{\setlength{\arraycolsep}{0pt} \setlength{\mathindent}{0pt}
\begin{eqnarray}
&&P(\Omega|\vg,\vb,I)=\kappa \int_0^\infty d\omega\,e^{-\omega}\,\times
\nonumber\\
&&\prod_{i=1}^{\ndet}\left\{\sum_{s_i=0}^{g_i}
\frac{(s_i+b_i)!}{b_i!s_i!(g_i-s_i)!}\,
\frac{(\omega\tau_i)^{g_i-s_i}(\td/\tb)^{s_i}}
     {(1+\td/\tb)^{s_i+b_i+1}}\right\},
\label{posterior_1}
\end{eqnarray}
}

\noindent where $\kappa$ is a normalization constant.  If instead of integrating
over the unknown background rates, we assume that $\tb$ is sufficiently
long that the observation of the $b_i$ determines the rates accurately,
we find instead the simpler approximate formula
{\setlength{\arraycolsep}{0pt} \setlength{\mathindent}{0pt}
\begin{eqnarray}
P(\Omega|\vg,\vb,I)\approx\kappa \int_0^\infty d\omega&&\,e^{-\omega}\,\times
\nonumber\\
&&\prod_{i=1}^{\ndet}[\omega\tau_i+(\td/\tb)b_i]^{g_i}.
\label{posterior_2}
\end{eqnarray}
}

In these equations, the $\tau_i(\Omega)$ represent our understanding of
the detector response.  We may estimate them by simple ray-tracing of
the mask pattern onto the detector, or they may be determined by Monte
Carlo simulations that account for physical effects (Compton scattering
by various elements of the spacecraft and instrument, finite detector
resolution, blurring due to the finite thickness of the coded aperture
mask, etc.) as well as purely geometric effects.

The procedure we have used to arrive at equations (1) and (2) may be
described as a joint fit of the background and event data, followed by
an integration over the unknown background Poisson means, which are
uninteresting parameters.  While this procedure may seem unfamiliar, it
is merely the Poisson version of the familiar Gaussian technique of
background subtraction.  In fact, if the data consisted of Gaussian
rather than Poisson deviates, with the likelihood proportional to
$e^{-\chi^2/2}$, we would be led directly to the usual formula for the
$\chi^2$ of a background-subtracted signal.  To press the analogy
further, the passage from eq. (\ref{posterior_1}) to eq.
(\ref{posterior_2}) is analogous to the neglect in Gaussian theory of
the contribution to the variance due to the uncertainty in the
measurement of the background.

Thus, this method is related to the ``$\chi^2$-fitting approach'' to
coded-aperture data analysis (Doty 1978) in the sense that, to within
an additive constant, $\chi^2$ is proportional to the log likelihood in
the Gaussian limit of many counts.  In fact, the likelihood method
constitutes a generalization of the $\chi^2$ method which is robust
even in the limit of low signal-to-noise.

We may use eqs. (\ref{posterior_1}) or (\ref{posterior_2}) (depending
on the background rate and on $\tb$) in two ways. We may estimate the
location of the transient on the sky by maximizing the probability
density with respect to $\Omega$, obtaining in effect the the maximum
likelihood estimate of the location.  We may also use the probability
density to obtain, say, a 68\% probability Bayesian confidence region
for the location of the transient.  We choose a grid of points that is
{\it uniform in an equal-area projection of the field of view} near
the estimated location of the transient, and calculate the probability
density at each point of the grid.  We choose $\kappa$ so that the sum
over the grid is normalized to 1, and find the value of the probability
density such that the sum of the probabilities of grid points that
exceed this value is 68\%.

\section{Simulations}
\subsection{Signal-to-Noise Study}

For illustrative and pedagogical purposes, we have simulated a highly
idealized coded-aperture instrument patterned after the HETE WXM.  This
detector consists of two crossed, one-dimensional position-sensitive
proportional counters (PSPCs), one each in the $x$ and $y$ directions.
The length of a PSPC bin is 0.1~cm. At a distance 18.73~cm above each
PSPC is a one-dimensional random array coded-aperture mask.  The
length of a mask element is 0.2~cm.  Note that the ``natural'' resolution
unit for this ideal detector is the ratio of the bin size to the height
of the mask above the PSPC, i.e. $\sim 18'$.

We model the $\tau_i$ using only geometric shadowing by the coded
aperture mask and the detector walls, and projection effects; no
physical effects (such as the ones alluded to in the previous section)
are included.  The encoded images obtained during the intervals $\td$
and $\tb$ are thus degraded by Poisson noise only.

\begin{figure*}[t]
\centering
\psbox{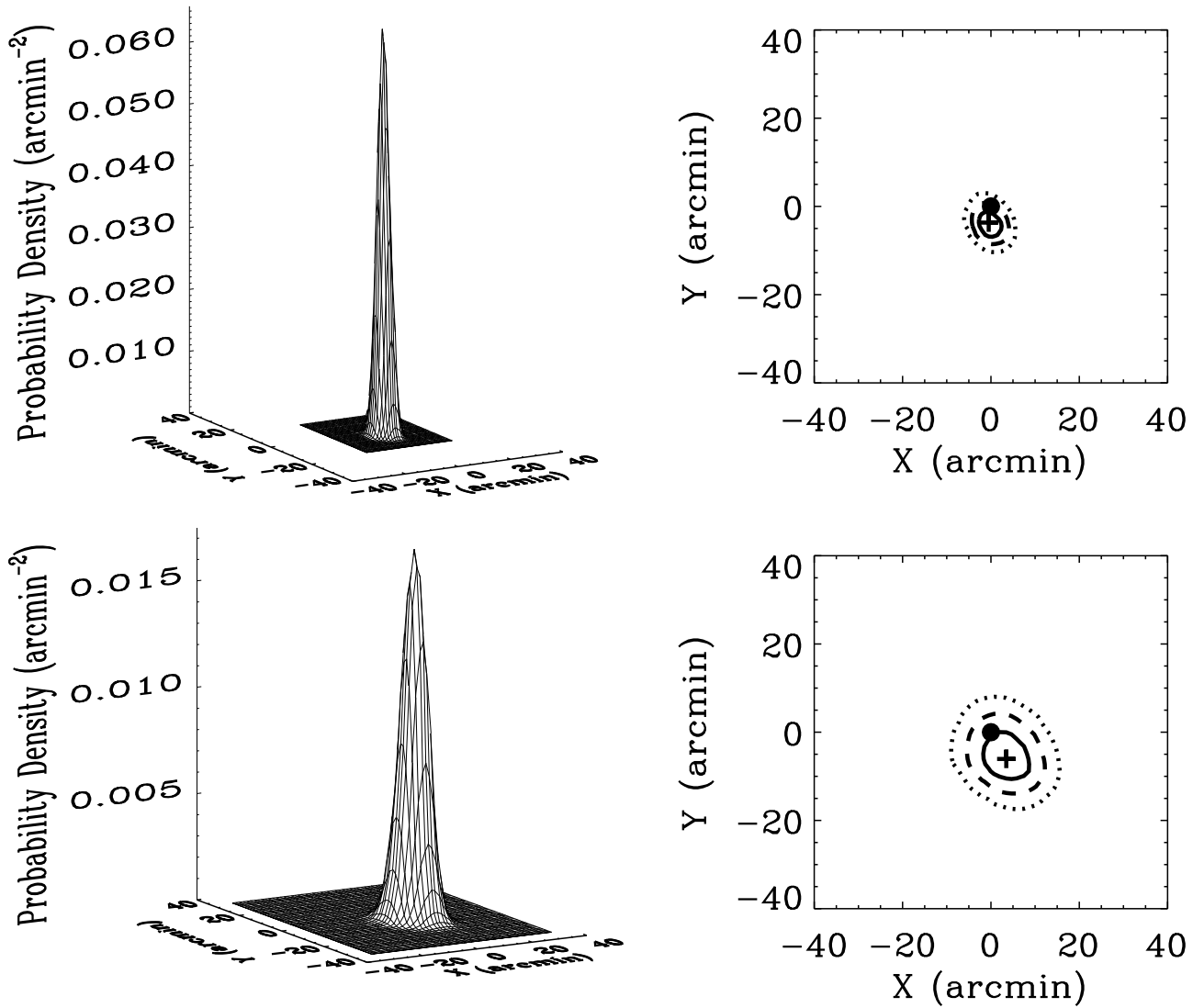}
\caption{Analysis of simulated data assuming a transient event from the
direction $\theta=30^\circ$, $\phi=45^\circ$.  The plots on the left
show the probability density, while the plots on the right show the 1-,
2-, and 3-$\sigma$ contours for the transient location, as well as the
true location (dot) and the maximum-likelihood point (cross).
The $X$ and $Y$ axes are cartesian coordinates in an equal-area
projection, shifted so as to place the true event location at the
origin. Top panel: S/N=10.  Bottom panel: S/N=5.}
\label{snr_10_5}
\end{figure*}

\begin{figure*}[t]
\centering
\psbox{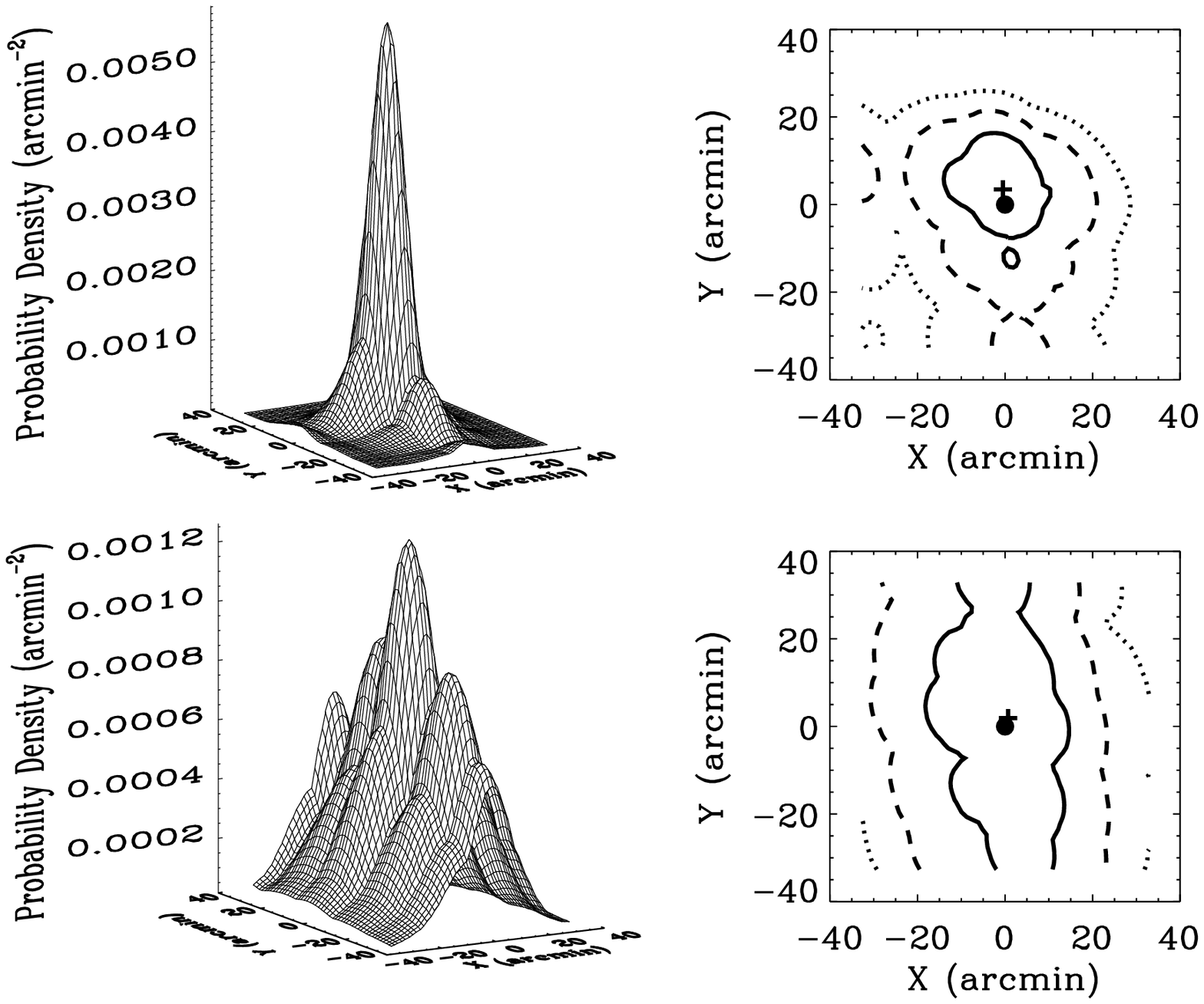}
\caption{Analysis of simulated data assuming a transient event from the
direction $\theta=30^\circ$, $\phi=45^\circ$.  The plots on the left
show the probability density, while the plots on the right show the 1-,
2-, and 3-$\sigma$ contours for the transient location, as well as the
true location (dot) and the maximum-likelihood point (cross).  The $X$
and $Y$ axes are cartesian coordinates in an equal-area projection,
shifted so as to place the true event location at the origin. Top
panel: S/N=10. Top panel: S/N=3.  Bottom panel: S/N=1.}
\label{snr_3_1}
\end{figure*}

Figures \ref{snr_10_5} and \ref{snr_3_1} show the result of analyzing
simulations at $(\theta=30^\circ,\phi=45^\circ)$, for signal-to-noise
ratios (S/N) of 10, 5, 3, and 1.  The probability density was
calculated using eq. (\ref{posterior_1}).  The solid, dashed, and
dotted contours represent 68.3\%, 95.5\%, and 99.7\% Bayesian
confidence regions, respectively.

The figures show that for high S/N, the angular resolution can be
considerably better than the ``natural'' resolution, while the
instrument's angular resolving power starts to ``melt down'' at about
S/N=3.  Note that the S/N is for the entire event, not just for the
trigger interval, so that even events discovered near or below a
3-$\sigma$ trigger threshold may be successfully located.

\subsection{Contour Calibration Study}

By construction, the contours chosen by this procedure have the
following property: If in simulations we choose locations from a
distribution of locations that is uniform in the FOV, and for each
location we simulate data and calculate a credible region, in the long
run the credible region will bracket the ``true'' location of the
simulated transient 68\% of the time.

We tested this property by simulating many transients events.  The
position of each event was drawn randomly from a uniform distribution
on the sky, limited to an angle from the detector normal of less than
20$^\circ$.  Each transient event was superposed on a background of 1
count~s$^{-1}$~bin$^{-1}$, and was assigned S/N=3
for the entire detector.  The background stretch lasted 100 sec for
each event.

Note that from the plots in Figure \ref{snr_3_1}, S/N=3 is about where
the linearity of the model in the position parameters begins to break
down, so that the posterior density looks significantly non-Gaussian. 
This is the regime where we would expect to lose faith in the
correctness of the calibration of contours obtained using the $\chi^2$
procedure of Lampton, Margon, and Bowyer (1976).

We simulated 1000 events, in each case calculating the 68.3\%, 95.5\%,
and 99.7\% contours and recording whether the contours included the
``true'' position.  The number of such ``hits'' is binomially
distributed, so that if the contours of probability value $p$ bracket
the true position $b$ times in $n$ attempts, we may use as a
statistical measure of the plausibility of this result the cumulative
distribution function $Q(b)\equiv\sum_{j=0}^b n!/j!(n-j)!\times
p^j(1-p)^{n-j}$.  $Q(b)$ should be approximately uniformly distributed
in the interval $[0,1]$, so the result will be plausible unless $Q(b)$
is found to be excessively close to either 0 or 1.

Out of 1000 simulated events, we found that the 68.3\% contour
bracketed the true event position 675 times ($Q=0.30$), the 95.5\%
contour did so 956 times ($Q=0.58$), and the 99.7\% contour did so 996
times ($Q=0.35$).  Thus the contours produced by the method appear to
be well-calibrated (as expected), in the sense that the long-term
frequency with which the contours bracket the true positions are indeed
consistent with their nominal probability values.

\section{Semi-Analytic Resolution Estimate}

We now introduce some analytical approximations in order to develop
some intuition regarding the method described in the previous section. 
These approximate estimates are helpful for optimization of detector
parameters, and also allow us to verify that our simulations of the
method, described in the previous section, behave as expected.

For concreteness, we specialize our analysis to coded-aperture systems
similar to the HETE WXM.  This is not a very serious restriction, since
other coded-aperture detectors (such as the ASM on the {\it Rossi}
X-Ray Timing Explorer) have fairly similar designs.  In any event,
generalization to other types of coded-aperture detectors is
straightforward.

The HETE WXM consists of two crossed, one-dimensional PSPCs, one each
in the $x$ and $y$ directions.  Above each PSPC is a one-dimensional
coded-aperture random array mask.

Let the PSPC bin the data in $\ndet$ position bins of length $\Delta$,
and let $\ndet\times\Delta\equiv\ldet$.  Also, assume that the mask
consists of $\nmask$ adjacent elements of length $l$, each of which may
be open or closed, and define $\nmask\times l\equiv\lmask$.  For HETE,
$\lmask>\ldet$, and $l>\Delta$, and we assume this is so in the present
analysis.  Let $h$ be the height of the mask above the PSPC.  We
further define the masks's open fraction $t$, which is such that
$t\nmask$ is equal to the number of open mask elements.  Finally, we
assume that the spatial resolution of the PSPC is described by a
smearing function $g_a(x-y)$, which is the probability that a photon
deposited at a position $y$ on the PSPC is recorded  with a position
$x$ in range $dx$.  It is assumed to have a width $a$.
Finally we define $\psi$ as the expected number of transient source
counts per bin assuming a $t=1$ mask, and $\eta$ as the expected number
of background counts per bin assuming a $t=1$ mask.

Assuming the direction of the transient is not too far from the
detector normal, we may evaluate the detector resolution by estimating
the width of the central peak of the probability density, in the form
given in eq. (\ref{posterior_2}). The angular variation of the
probability density may be traced to the variation with the assumed
transient source position of the shadows of mask element edges on the
detector.  The angular resolution thus increases with increasing number
of edge shadows on the detector, as well as with increasing sharpness
of the edge shadows. Guided by this observation we have shown that the
detector angular resolution $\delta\theta$ is approximately given by
%
\begin{eqnarray}
\delta\theta&\approx&
S_a^{-1} \bigg\{
2 F(\theta)\left(\frac{h}{\Delta}\right)^2
\left(\frac{\ldet}{l}\right)
\left(\frac{\psi^2}{\eta}\right)\times\nonumber\\
&&\hspace{3.5cm}\frac{1-t}{1+\frac{\psi}{2\eta}\frac{1}{t}}
\bigg\}^{-\frac{1}{2}},
\label{thegoodies}
\end{eqnarray}
where
\begin{equation}
S_a\equiv\int_{-\Delta/2}^{+\Delta/2}dx\,g_a(x),
\label{smearing}
\end{equation}
and where $F(\theta)$ is the fraction of the detector that is not
shadowed by the detector walls for a transient source at $\theta$. For
the 1000 simulations discussed in the previous section, we find that
this expression, translated into a confidence region, is in good
agreement (10\%-20\%) with the actual confidence region sizes.

Equation (\ref{thegoodies}) may be readily interpreted as follows:

As is intuitively reasonable, $\delta\theta\propto\Delta/h$. Thus the
angular resolution is naturally expressed in units of the angle
subtended at the mask by a detector bin size.  It follows that it is
desirable to design the coded-aperture system with as small a
$\Delta/h$ as possible.  Of course, increasing $h$ reduces the field of
view and increases the penalty imposed by the factor $F(\theta)^{-1/2}$
for off-axis events.

Furthermore, $\delta\theta\propto(\ldet/l)^{-1/2}$, which accounts for
the ``tiling'' of the detector by the mask elements --- this factor in
effect counts the number of mask element edges that cast their shadows
on the detector.  Clearly, it places a premium on as small an $l$ as
possible, that is, $l=\Delta$.  Thus it appears that ``oversampling''
of the mask by the detector is not beneficial.

When $\psi\gg\eta$, we have $\delta\theta\propto 1/\psi^{1/2}$, while
when $\psi\ll\eta$, we have $\delta\theta\propto\eta^{1/2}/\psi$. 
Thus in either case, $\delta\theta\propto 1/$SNR.

\begin{figure*}[t]
\centering
\psbox[ysize=0.25#2]{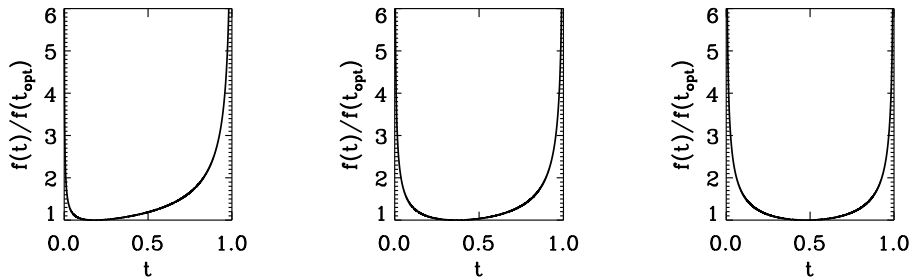}
\caption{Angular resolution as a function of transmission fraction $t$,
for $\psi/\eta$=0.1, 1, and 10. The function $f(t)$ is given by 
$[(1+\psi/2\eta t)/(1-t)]^{1/2}$.  The quantity $\topt$ is the
optimum transmission fraction for the given signal-to-background.
}
\label{transmission}
\end{figure*}

The dependence of $\delta\theta$ on $t$ is relatively simple.
There is an
``optimal'' $\topt$ that minimizes $\delta\theta$, given by
\begin{equation}
\topt=-\frac{\psi}{2\eta}+\frac{1}{2}\left[\left(\frac{\psi}{\eta}\right)^2
+2\frac{\psi}{\eta}\right]^\frac{1}{2}.
\label{optimum_formula}
\end{equation}
\begin{figure}[h]
\centering
\psbox[ysize=0.8#2]{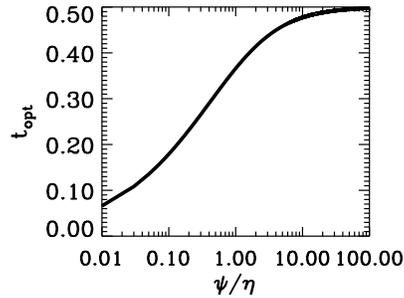}
\caption{Optimum transmission fraction as a function of
signal-to-background.}
\label{optimum_trans}
\end{figure}
The behavior of $\topt$ as a function of $\psi/\eta$ is shown in Figure
\ref{optimum_trans}.  The form of $\topt$ is different from the one
found by previous authors (Fenimore 1978, in't Zand, Heise, \& Jager
1994), although the functional dependence is still on the
signal-to-background.  The difference is ascribable to the difference
in the quantities optimized:  the previous work optimized the
signal-to-noise in the resolution element corresponding to the source,
whereas in this work we optimize the angular resolution directly.

Figure \ref{transmission} shows the $t$ dependence of $\delta\theta$
for signal-to-background ratios of 0.1, 1, and 10.  It is clear from
this figure that precisely optimizing $t$ is not critical.  The
$t$-dependence of the resolution is a very forgiving function, with
relatively large differences from $\topt$ costing relatively little in
resolution.

Finally, the factor $S_a$ accounts for the smearing by the detector of the
otherwise perfectly sharp shadows of the mask element edges.  It is the
``migration'' with assumed source position of these shadows across the
PSPC bins that gives the probability density its sensitivity to the
transient location, and which makes possible resolutions
$\delta\theta<\Delta/h$.  Naturally, the linear resolution of the PSPC
degrades the angular resolution of the instrument, by a factor given by
$S_a$.  Note that $S_a\rightarrow 1$ as $a\rightarrow 0$.

\section{Conclusions}

The likelihood function approach to coded-aperture data analysis is a
very powerful method that makes maximal use of the information borne by
the data.  The method produces not only point estimates of source
position, but also credible regions that are well-calibrated in the
sense that in the long run, a 68\% region brackets the true location in
68\% of simulations.  This remains the case even in the low
signal-to-noise case, where linear methods lose their statistical
reliability.  The method can produce angular resolution in excess of the
``natural'' resolution of the detector, for high signal-to-noise ratios.

In addition, a semi-analytic estimate of the angular resolution gives a
result that is intuitively plausible and provides useful guidance for
coded-aperture instrument design.  In particular it provides a novel
criterion for assessing the impact of the choice of transmission
fraction of the coded-aperture mask, and indicates that it is not
beneficial to design a position-sensitive detector that ``oversamples''
the mask pattern.

\section*{Acknowledgements}

We thank Tom Loredo for clarifying the calibration properties of
Bayesian credible regions, and John Doty for pointing out an error in
an early version of the paper.

\section*{References}

\re
Doty~J.~P. 1978, in X-Ray Instrumentation in Astronomy II, ed.
L.~Golub, Proc SPIE, 982, 164

\re
Fenimore~E.~E. 1978, Appl.~Opt., 17, 3562

\re
in't~Zand~J.~J.~M., Heise~J., and Jager,~R., 1994, A\&A, 288, 665

\re
Lampton~M., Margon~B., and Bowyer~S., 1976, ApJ, 208, 177

\re
Loredo~T.~J. 1992, in Statistical Challenges in Modern Astronomy, ed.
E. Feigelson and G. Babu (New York: Springer-Verlag), p. 275

\re
Ricker~G.~R. 1997, these proceedings

\re
Sims~M., Turner~M.J.L., and Willingale~R. 1980, Space Sci. Instrum., 5, 109

\re
Skinner~G.~K., and Nottingham M.~R. 1993, Nucl. Instrum. Methods Phys.
Res., A333, 540

\re
Willingale~R., Sims~M., and Turner~M.J.L. 1984, Nucl. Instrum. Methods 
Phys. Res., 221, 60

\end{document}